# Correlated Effects of Relative Size and Depth in the Perceptual Organization of Multiple Figure-Ground Configurations

Birgitta Dresp-Langley
Centre National de la Recherche Scientifique (CNRS)
UMR 7357 ICube Lab Strasbourg University
Strasbourg, France
e-mail: birgitta.dresp@unistra.fr

Adam Reeves
Department of Psychology
Northeastern University
Boston, USA
e-mail: a.reeves@northeastern.edu

*Abstract*—The neural networks of the human visual brain derive representations of three-dimensional structure from specific two-dimensional image cues. Neural models backed by psychophysical data predict how local differences in either luminance contrast or physical size of local boundaries in 2D images may determine the perception of 3D properties. Predictions relative to the role of color in this process do not follow from any of the current models. To further clarify the potential contribution of color to perceptual organization, image configurations with multiple surface representations where the relative physical size of local boundaries between contrast regions was held constant were submitted to perceptual judgments of relative size and relative depth. The only potential cues available in the images were generated by the specific local combinations of color and luminance contrast. It is shown that response probabilities for subjective depth and subjective size are systematically and consistently determined by local surface colors and their immediate backgrounds. There is a statistically significant correlation between subjective depth and subjective size, and a color specific effect on both dependent variables. This effect depends on the polarity of the immediate surround of the reference surface rather than on local center-surround contrast intensity. It is suggested that the underlying neural mechanisms selectively exploit specific color and background cues to enable intrinsically coherent perceptual organization of the otherwise highly ambiguous image input.

<u>*Keywords*</u>: *colour; local contrast; background polarity; multiple surfaces; figure-ground; subjective size; relative depth.*

BACKGROUND

Since Leonardo da Vinci [1] local luminance contrast has been pointed out as a perceptual cue to three-dimensional properties of objects depicted in the Euclidean plane. Contemporary neural models and psychophysical data predict that contrast variations across image parts directly determine which parts of a planar image will be seen as "nearer" or "further away from" the human observer [1] - [10]. Previous studies on functional aspects of mechanisms for depth perception from neural computation of local image contrast properties have not yet fully explored all the complex interactions between color, luminance, and general background field intensities. In the absence of other spatial cues to depth, it appears that specific colors in combination with specific contrast intensities may produce more powerful 3D effects than others, as suggested by results on perceptual figure-ground organization, for example [10] [1] - [14]. Moreover, variations in brightness or luminance displayed across two or more different surface layers in complex multiple-surface configurations may alter these perceptual effects significantly [7] [14] [15], or even reverse them [16] [17]. This study was designed to explore some of such possible interactions more systematically. Complex image configurations with carefully controlled physical variations in local color, luminance, general background intensity, and constant spatial parameters were generated for this purpose. The local physical size of the test and reference surfaces submitted to perceptual judgments was not varied across comparisons. Center-surround surface combinations within image configurations were displayed on a high resolution monitor in a computer controlled psychophysical study with human subjects completing four Two-Alternative spatial Forced Choice (2AFC) judgment tasks. The subjects had to judge which of two comparison surfaces in the configurations appeared "bigger" (task 1) or "nearer" (task 3), and which of all the possible reference surfaces in a given configuration appeared "the biggest" (task 3) or "the nearest" (task 4). Materials and methods used to generate the image configurations for this study, some of the characteristics of the study population, and the experimental task procedures are explained here below. Results, discussion of implications for our current understanding of the perceptual organization of ambiguous image input, and a short conclusion are provided subsequently.

MATERIALS AND METHODS

All image configurations were computer generated and displayed on a high resolution color monitor (EIZO COLOR EDGE CG 275W, 2560x1440 pixel resolution) connected to a DELL computer equipped with a high performance graphics card (NVIDIA). Color and luminance calibration of the RGB channels of the monitor was performed using the appropriate Color Navigator self-calibration software, which was delivered with the screen and runs under Windows 7. RGB values here correspond to ADOBE RGB. All luminance levels were cross-checked with an external photometer (OPTICAL, Cambridge Research Systems). RGB coordinates, luminance parameters (cd/m$^2$), and color coordinates (X, Y, Z) of the different reference surfaces in the image configurations from this study are given in Table 1.

The size of each of the square surfaces in the center of each of the twelve local configurations in the images was 160x160 pixels and the size of each of the square surrounds was 400x400 pixels. The twelve local configurations were equally spaced, with 50 pixels between their surrounds, along the horizontal and vertical dimensions. They were displayed centrally on the dark and light general background of the 2560x1440 pixel screen. The size of a single pixel on the screen is 0.023 cm. Grey, red, and blue-green center squares on their light and dark immediate surrounds were presented in pairs, as shown in Figure 1. Their position (left, right) in a pair was counterbalanced between trials and subjects. Presentation on light and dark general backgrounds was also counterbalanced between trials and subjects. The subject pool consisted of mostly undergraduate medical students, with normal or corrected-to-normal vision. All of them were naïve to the purpose of the experiment and run in individual sessions. They were comfortably seated in a semi-dark room, in front of the EIZO monitor at a viewing distance of about 1 meter. Each individual received the same standard instructions for the psychophysical tasks. In one task, the subject had to decide which of the two central squares in a paired configuration (*paired comparison*) appeared "bigger". In the other task, the subject had to pick the central square from all of the twelve configurations that appeared the "biggest" (*single pick*). In another task, the subject was instructed to judge which of the two central squares in a paired configuration (*paired comparison*) appeared to be "nearer" to them, and in a fourth task he/she had to pick the central square from all of the twelve configurations that appeared the "nearest" (*single pick*) to the observer.

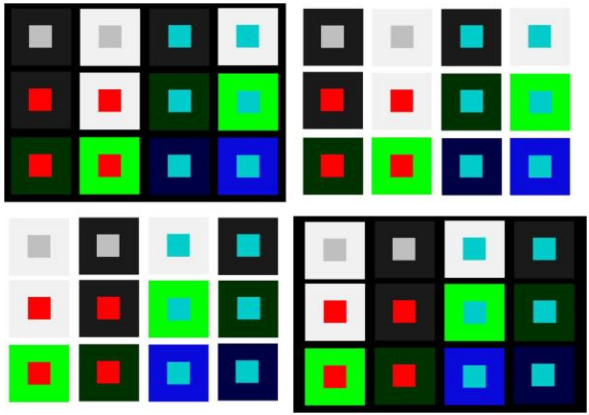

Figure 1. Multiple figure-ground configurations on dark and light general backgrounds.

The twelve local configurations shown in Figure 1 produce subjective differences in the relative size and depth of the centrally displayed squares. Grey, red, and blue-green center squares displayed on dark and light surrounds were paired for the relative psychophysical judgments. Trials were sequenced in counterbalanced sessions producing eight psychophysical judgments for each *paired comparison* and *single pick* task, general background condition, and subject. Therefore, a total of 80 data was generated for each of the four tasks and for each of the two general background conditions

The subjects who participated in this study were adult volunteers naïve to the purpose of the study. We selected seven men and three women with normal or corrected-to-normal vision. The experiments were non-invasive and conducted in accordance with the Declaration of Helsinki (1964) and with full approval of the corresponding author's host institution's (CNRS) ethical standards committee. Informed consent was obtained from each of the participants.

TABLE I. COLOR AND LUMINANCE

| Reference Surface | Image Luminance (*L*) And Color Coordinates | | | | | | |
|---|---|---|---|---|---|---|---|
| | *R* | *G* | *B* | $L$ $(cd/m^2)$ | X | Y | Z |
| Grey Center | 190 | 190 | 190 | 58.6 | 49.8 | 52.3 | 57.0 |
| Red Center | 255 | 0 | 0 | 35.8 | 57.7 | 29.7 | 2.7 |
| Blue Center | 0 | 205 | 205 | 52.3 | 23.1 | 43.5 | 65.7 |
| Dark-Grey Surround | 25 | 25 | 25 | 2.0 | 0.6 | 0.6 | 0.6 |
| Light-Grey Surround | 240 | 240 | 240 | 95.3 | 83.2 | 87.5 | 95.3 |
| Dark-Green Surround | 0 | 50 | 0 | 2 | 0.5 | 1.7 | 0.2 |
| Light-Green Surround | 0 | 255 | 0 | 78.5 | 18.5 | 62.7 | 7.1 |
| Dark-Blue Surround | 0 | 0 | 70 | 0.5 | 1.1 | 0.4 | 5.8 |
| Light-Blue Surround | 10 | 10 | 220 | 5 | 13.7 | 5.52 | 71.6 |
| Dark General Background | 0 | 0 | 0 | 0.5 | 0 | 0 | 0 |
| Light General Background | 255 | 255 | 255 | 120.0 | 13.7 | 5.52 | 71.6 |

RESULTS AND DISCUSSION

The response probabilities (*p*) from the two paired comparison tasks (task 1, task 3) were calculated for each of the twelve local center-surround configurations in the order in which they are displayed in the first of the four general display-panels shown in Figure 1. A *p* of 1 would correspond to the case where a local configuration of a given pair produces a total number of 80 observed/80 possible responses for "bigger" or for "nearer". In this case, the *p* associated with the other configuration from that pair would be 0. In the case a given pair produces random perceptual responses for "bigger" or for "nearer", the response probability associated with each of the two paired configurations would be 0.50. In a first analysis, the twelve configurations were sorted as a function of the magnitude of the response probabilities they produced for "bigger" and "nearer" and plotted in ascending order for each of the two "general background intensity" conditions. These plots are shown in Figure 2. The two graphs reveal consistent *p* distributions for "bigger" and nearer" ranging from 0.10 to 0.90 in each of the two general background conditions.

The response probability distributions from the paired comparison tasks were submitted to statistical correlation analyses (Pearson's product moment), returning statistically significant correlation coefficients (*P*), with 0.98 (p<.001) for "bigger" and "nearer" in the "dark general background" condition, and 0.99 (p<.001) for the probability distributions for "bigger" and "nearer" in the "light general background" condition. These analyses show that the center-surround configurations produced a wide range of significantly correlated perceptual differences in relative size and depth of their local center surfaces.

In a second analysis, the configurations were sorted as a function of their local contrast intensity. The luminance contrasts (*LumC*) are expressed here in terms of Weber Ratios, which are calculated using

$$LumC = Lum_{center} - Lum_{surround} / Lum_{surround} \quad (1)$$

The response probabilities for "bigger" and "nearer" were then plotted as a function of the twelve different Weber contrasts of the configurations and the two general background conditions, shown in Figure 2. Graphs in the top panel show significantly correlated magnitudes of *p* for "bigger" and "nearer" produced by the twelve configurations on the two general backgrounds, plotted in ascending order. The graphs in the middle panel show *p* distributions as a function of the luminance contrast intensity (Weber ratios) of the twelve configurations and the general background conditions. The graphs in the bottom panel show *p* as a function of the local color contrast of the configurations with positive (+) Weber contrasts, which produced greater magnitudes of p for "bigger" and "nearer" in the paired comparison tasks.

The data show that no simple function links the *p* for relative size and depth to the luminance contrast of the local configurations. There is a systematic effect of the general background condition on all the *p*: the lighter general background produced systematically stronger response probabilities for "bigger" and "nearer". The configurations with the positive local contrast signs all produced greater magnitudes of *p* in comparison with their negative-contrast-sign pairs, however, the configurations with the strongest positive contrasts did not produce the highest response probabilities, neither for "bigger" (relative size), nor for "nearer" (relative depth) in the paired comparison tasks.

This is clarified further by the graphs shown in the panel at the bottom of Figure 2, where *p* for "bigger" and "nearer" are shown as a function of the local color contrast of the configurations which produced the stronger *p* magnitudes, and as a function of the general background condition. The highest *p* for "bigger" (relative size) and "nearer" (relative depth) are produced by the RED central squares on the dark-grey surround displayed on the light general background, and by the GREY central squares on the dark grey surrounds displayed on the general background. The BLUE central squares on dark surrounds produced noticeably lower p for "bigger" and "nearer" in comparison with the RED centers, yet, the blue on dark surrounds has a much stronger luminance contrast (25.5) than the red on dark surrounds (16.9). For the blue centers on dark surrounds we observe the strongest effect of general display background condition: the *p* for "bigger" and "nearer" are well above a certain positive probability threshold (>=0.75) for the blue-on-dark configurations displayed on a light general background, but approach the chance level (~0.50) in the condition where they were displayed on a dark general background.

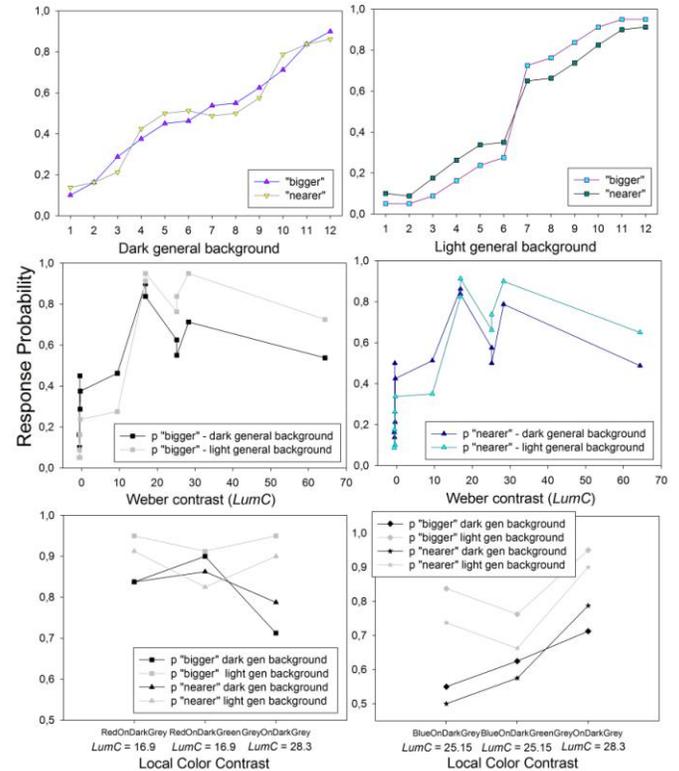

Figure 2. Probability distributions for relative size and relative depth

The *p* distributions from the two *single pick* tasks were also plotted as a function of the contrast intensity of the twelve local configurations and the general display background condition. These results, shown in Figure 3, consistently indicate that the highest response probabilities for "biggest" and "nearest" are produced by RED centers on dark GREEN or GREY local surrounds. This result is consistent with earlier observations [9] and further highlights the hitherto not shown dependency of this selective color effect on physical parameters relative to the immediate and the general background intensities. These results highlight complex interactions between color, local luminance contrast and global display background in the production of perceptual effects of subjective relative size and depth. Some of them, but not all, are predicted by current neural theories [6] [10] [15] [17] [18].

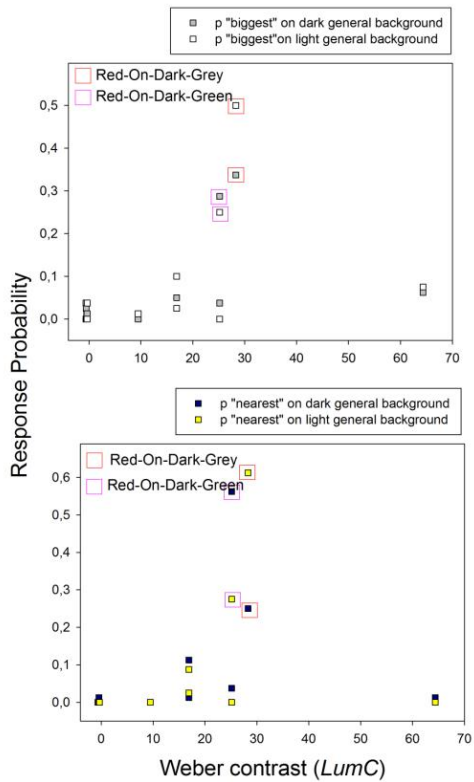

Figure 3. Probability distributions for "biggest" and "nearest" as a function of local Weber Ratios and general background.

CONCLUSIONS

The subjective relative size of surface boundaries is significantly correlated with subjective depth, generating a perceptual 2D cue to 3D structure functionally equivalent to the "real", physically grounded, monocular depth cue. Although not explicitly predicted by any of the neural models, this observation supports a specific class of computational models of relative surface depth from ambiguous contrast input generated by 2D contrast surfaces and their boundaries [7] [11] [15] [17]. The perceptual judgments from this study support the idea that the human brain is capable of producing functional perceptual representations of figure and ground by using a multitude of different local and global cues in the 2D image. Ambiguous image input is dealt with by effectively exploiting whatever cue to structure is available in the display. Absence of a specific physically grounded depth cue may be compensated for by perceptually generated cues that allow the brain to compute coherent depth representations on the basis of global perceptual sensation rather than merely the direct or strictly local visual processing of an existing stimulus parameter such as a physical difference in the size of 2D surface boundaries, for example. Also, the way in which contrast is computed to achieve perceptual 3D structure reaches well beyond local processing. As shown here, the lighter general backgrounds of the configurations, resulting in image representations with more than two 2D surface layers, systematically produced stronger subjective depth effects, irrespective of the local color or contrast of the reference surfaces and their immediate surrounds. This has potentially important implications for the development of effective visual interface technology for image-guided systems designed to assist human operators in precision tasks [16]. The results from this study are consistent with previous findings that the color red is the most likely to produce depth effects in simple figure-ground displays with only two 2D surface representations [3] [9]. Red surface color on an achromatic background, for example, possesses a clear competitive advantage over other colors such as green [9] or blue [12] in the likelihood to be perceived as closer to the human observer. As shown here, when more than two surface layers are present in an image configuration, the advantage of the surface color red for perceptual organization appears to depend on the contrast polarities of all the image regions surrounding the reference surface, not on the local reference-surround luminance contrast. This result is new and may seem surprising, yet, it is fully consistent with experimental evidence from other studies showing that many different cues may cooperate adaptively and non-locally in figure–ground segregation from 2D cues [19]. Physiologically inspired model approaches which simulate how figure–ground segregation may be computed by neural mechanisms "beyond the classic receptive field", involving long-range feedback interactions between cells with increasingly larger receptive fields in higher visual cortical areas beyond V1, V2, or even V4 [15] [19] [20], are in principle suitable to account for the non-local processing of figure-ground. However, it is not clear how these models would account for selective color effects, as those shown here. It is possible that these effects may be linked to psychological effects of selective attention to specific colors and/or color cognition in a more general sense [21] [22]. These are still poorly understood and warrant to be investigated further. Although color detection is hardwired in the brain and machines can learn to detect colors even more reliably than the human vision [23] [24], early color vision is not impenetrable [25] and may be subject to top-down modulation.